\documentclass[11pt]{iopart}
\usepackage{epsfig}
\begin{document}
\title{Towards QCD thermodynamics using exact chiral symmetry on lattice }
\author{ R V Gavai and Sayantan Sharma }
\address{Department of Theoretical Physics, Tata Institute of Fundamental
         Research,\\ Homi Bhabha Road, Mumbai 400005, India.}
\ead{gavai@tifr.res.in,ssharma@theory.tifr.res.in }
\begin{abstract}
The thermodynamics of massless ideal gas of overlap quarks has been
investigated numerically for both zero and nonzero baryon
chemical potential $\mu$. While the parameter $M$ has been shown to be irrelevant in the continuum limit, it is shown numerically that the continuum limit can be reached with relatively
coarser lattices for certain ranges of $M$. Numerical limitation of the
existing method of introduction of chemical potential in the overlap formalism is discussed. We have also studied the energy density of free domain wall fermions in the absence of $\mu$  and estimated the extent of lattice in the fifth dimension $L_5$ for which the overlap results are recovered. Interestingly, this value of $L_5$ is also minimum for the same range of $M$ found in the overlap case. 
\end{abstract}

\maketitle
\section{Introduction}

     The phase diagram of Quantum Chromodynamics(QCD) has been an important subject of study in the recent years and lattice gauge theory has emerged as a major tool for studying it. 
     It is believed that spontaneously broken chiral symmetry is restored at high temperatures so it is very important to study the QCD thermodynamics with fermions having exact chiral symmetry on lattice. The most popular among such fermions are the overlap and the domain wall fermions. The overlap operator is highly non-local, involving inversion and square root of Dirac matrices,  making it computationally very expensive to simulate. The domain wall fermions on the other hand are necessarily defined on a 5-D lattice so their computational cost  is more than the standard 4-D fermions as the Wilson and staggerred, but is still less than the overlap. For the above mentioned practical reasons, we investigate numerically whether the irrelevant parameter $M$ in such fermion operators can be tuned optimally to recover the continuum values of various thermodynamic quantities for free fermions with the smallest possible lattice sizes. This would give us an estimate of the optimum lattice parameters for full QCD computations with such fermions. In section 2, we compute the energy density of overlap and domain wall fermions in absence of chemical potential. It was shown\cite{neu} that the domain wall operator reduces to the overlap when the extent of lattice in the fifth dimension, $L_5\rightarrow\infty$. We try to find out at what  $L_5$, the overlap results are obtained. In the subsequent sections the energy density in the presence of chemical potential $\mu$ and the susceptibility expressions are computed and their approach to the continuum for different values of M are studied.
\section{Energy density in absence of chemical potential}
\subsection{Overlap fermions}
The overlap Dirac operator\cite{NeuNar} for massless quarks defined on a $N^3\times N_T$ lattice with spacing a and $a_4$ in the spatial and temporal directions respectively, is given by,
\begin{equation}
 D_{ov}=1+\gamma_5 sgn(\gamma_5 D_W)~,~
\end{equation}
where sgn denotes the sign function and 
\begin{eqnarray}
\label{eqn:Dwil}
\nonumber
D_W(x,y)& =& 
-M~\delta_{x,y}-\frac{a}{a_4}[\delta_{x-\hat{4},y}
\frac{1+\gamma_{4}}{2} +\frac{1-\gamma_{4}}{2}\delta_{x+\hat{4},y}]\\ 
&-&\sum_{i=1}^{3}
[\delta_{x-\hat{i},y}\frac{1+\gamma_{i}}{2} 
+\frac{1-\gamma_{i}}{2}\delta_{x+\hat{i},y}]+[3+\frac{a}{a_4}]\delta_{x,y}.
\end{eqnarray}
 The sign function of the matrix $\gamma_5 D_W$  can be written as the sign of its eigenvalues\cite{gatt}. These eigenvalues are positive and greater than zero in this case so the sign function is well-defined. The overlap operator hence can be diagonalised in the momentum space in terms of the variables
\begin{equation}
\label{eqn:def}
 h_i=-\sin a p_i~,~ h_4= -\frac{a}{a_4}\sin a_4 p_4~,~ h_5=M-\sum_{i=1}^{3}(1-\cos a p_i)-\frac{a}{a_4}(1- \cos a_4 p_4)~.~
\end{equation}
The energy density can be obtained as
\begin{eqnarray}
\label{eqn:ed}
\nonumber
 \epsilon &=& \frac{T^2}{V}
     \left.\frac{\partial\ln Z(V,T)}{\partial T}\right|_ V
=-\frac{1}{N^3 a^3 N_T}\left(\frac{\partial
   ln \det D_{ov}}{\partial a_4}\right)_{a}
 \\
 &=& \frac{2}{N^3 a^4 N_T} \sum_{p_j,p_4}\frac{h^2 (1-\cos a p_4)-h_5 h_4  \sin a p_4 }{h^2 (h^2+h_5^2)}(\sqrt{h^2+h_5^2}+h_5)~,~
\end{eqnarray}
where  $a=a_4$  has been chosen after performing the $a_4$-derivative. The same can be computed numerically keeping the aspect ratio $\zeta=N/N_T$ fixed. The lattice expression so obtained was fitted to the ansatz $A+B/N_T^4$ for a range of values of M. The higher order $N_T$ dependent correction terms in the energy density expression due to finite size effects manifest themselves as an effective coefficient $B(M)$. We seek those values of M for which $B(M)$ approaches $7 \pi^2/60$, its value in the continuum limit. The zero temperature part A, was subtracted from the energy density. The resultant $\epsilon$ divided by continuum value $\epsilon_{SB}$ was plotted as a function of $N_T$. From figure 1(a) we  observe that the M-dependence is quite pronounced and for $1.55<M<1.60$ the  continuum limit is reached for the smallest possible lattice size($N_T\sim 12$). For the oft-favoured choice of $M=1$, the convergence to the continuum is seen to be very slow. The energy density can be evaluated analytically by converting the sum over Matsubara frequencies $a p_4=\omega=(2m+1)\pi/N_T$ where m are integers, to contour integrals\cite{gavai}.  It has been shown that no M-dependence of the energy density exist in the continuum\cite{gavai} but for observing that we need lattice sizes much greater than $N_T=32$.
\begin{figure}
\begin{center}
 \includegraphics[scale=0.58]{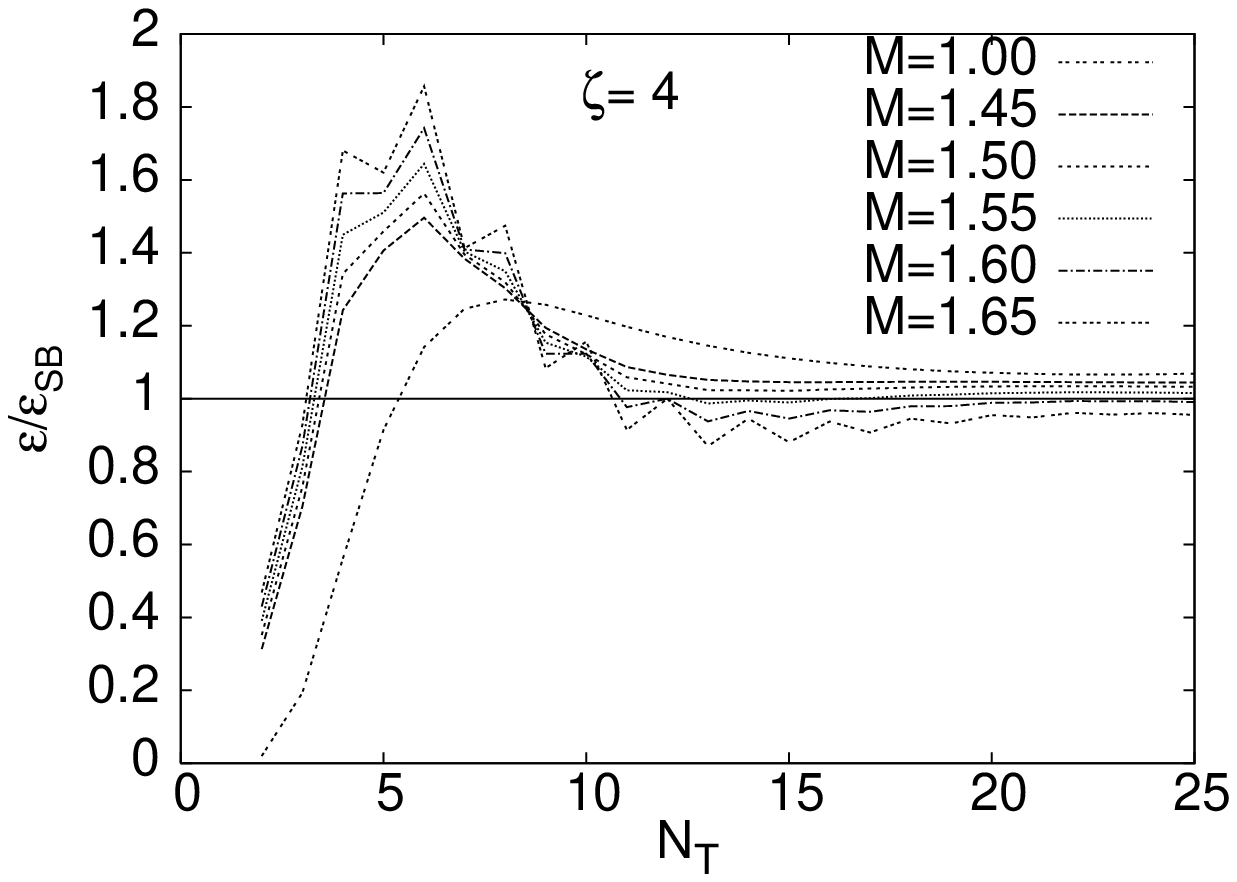}
\includegraphics[scale=0.62]{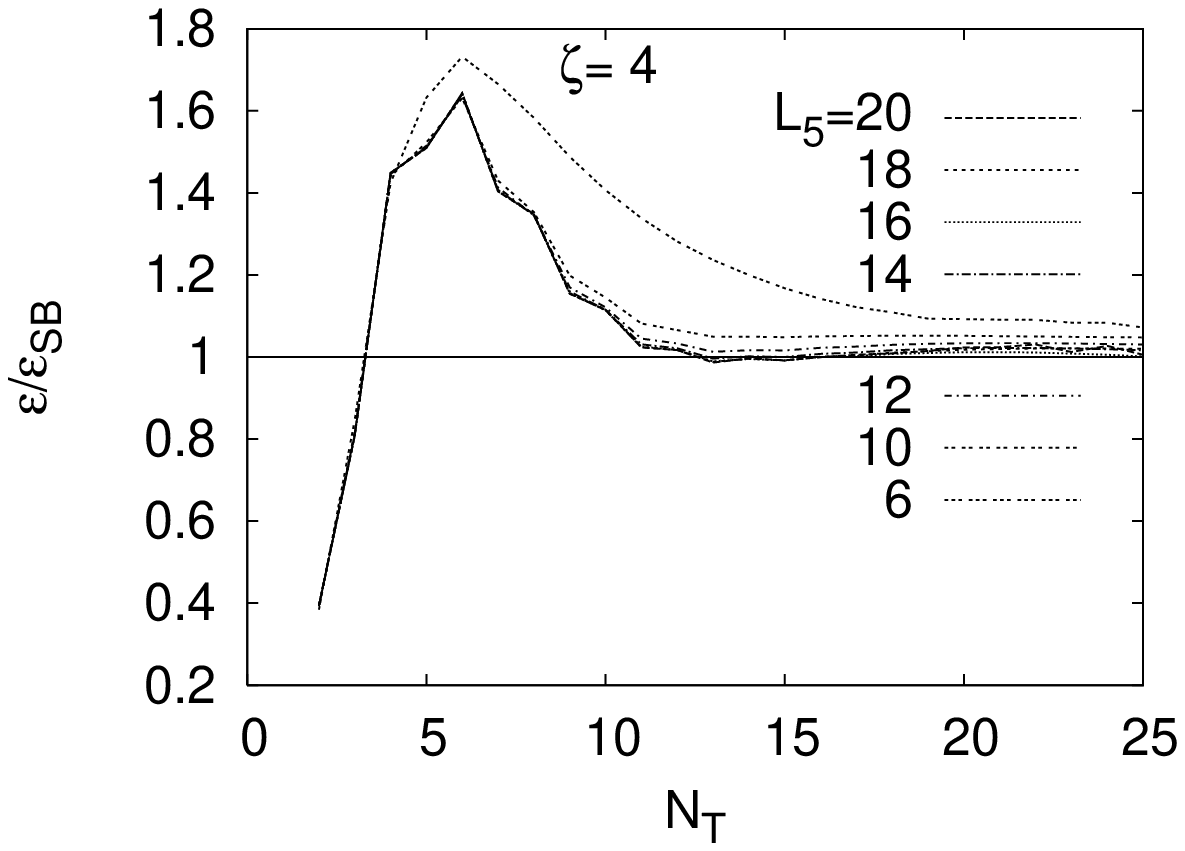}
\label{fig.1}
\caption{(a)The variation of $\epsilon/\epsilon_{SB}$ with M for fixed $\zeta=4$ for overlap fermions and (b)The domain wall fermion results matches with the overlap for $L_5\geq14$ with $\zeta=4$ and $M=1.55$. }
\end{center}
\end{figure}
\subsection{Domain wall fermions}
The domain wall fermions\cite{kap} are defined on a 5-D Euclidian lattice in presence of a  domain wall of height M. 
Taking  the lattice spacing in the fifth direction $a_5\rightarrow 0$ and the number of sites $N_5\rightarrow \infty$ for fixed $L_5=N_5a_5$, the effective domain wall operator is
\begin{equation}
 D_{DW}=(1+\frac{m_q}{2M})+(1-\frac{m_q}{2M})\gamma^5\tanh(\frac{L_5}{2} \gamma^5 D_W)~. ~
\end{equation}
Here $m_q$ is the bare quark mass and $L_5$ is in lattice units. 
In the momentum space, using $h_1$, $h_2$, $h_3$, $h_4$ and $h_5$ as given in (\ref{eqn:def}), the energy density for massless quarks($m_q=0$) is given by 
\begin{equation*}
\varepsilon a^4=\sum_{p_j,p_4}\frac{4 \sinh[\frac{sL_5}{2}]( \left(-h_4 h_5 \alpha +h^2 \gamma \right)  \cosh[\frac{sL_5}{2}]+\left(h_4 h_5 \alpha +(h_5^2+s^2)\gamma+2h_5 s^2 p)\right)}{sN^3N_T(h^2+\left(s^2+ h_5^2\right) \cosh[2sL_5]} 
\end{equation*}
\begin{equation}
 \frac{\cosh(\frac{3 s L_5 }{2})-2 s (h_5^2 p+h_5\gamma +(h^2 p+h_4 \alpha +2 h_5(h_5 p+\gamma))  \cosh[ sL_5]) \sinh[\frac{s L_5}{2}])}{-2 h_5 s \sinh[2 s L_5])}~,~
\end{equation}

where 
\begin{equation}
 \alpha=-h_4~ ,~\gamma=1-\cos((2m+1)\pi/N_T)~,
\end{equation}
\begin{equation}
h^2=h_1^2+h_2^2+h_3^2+h_4^2~,~ s^2=h^2+h_5^2 ~,~
\end{equation}

\begin{equation}
 p=\frac{(-\sin^2 ap_4+h_5 (1- \cos ap_4))(-\tanh \frac{ L_5 s}{2}+\frac{L_5 s}{2} sech ^2 \frac{L_5 s}{2})}{s^2\tanh \frac{L_5 s}{2}}~.~
\end{equation}

 Following the same fitting procedure mentioned in the previous section and choosing $M=1.55$ the $N_T$ dependent part of energy density $\epsilon$  was plotted for various values of $L_5$. 
Clearly from figure 1(b) we observe that for $L_5 \geq 14 $, the energy density converges to the continuum  case at $N_T\geq12$. $L_5\sim14$ is thus the optimum lattice size in the fifth dimension to converge to the overlap results.\\
\section{Non-zero chemical potential}
In the presence of chemical potential $\hat\mu=\mu a$ there is a critical value of $\hat\mu_c$ beyond which the sign function is not always defined\cite{gavai}. Hence we work with $\hat\mu<\hat\mu_c$,  where the argument of sign function is positive definite and the calculation follows analogously as in section $2.1$. The chemical potential can be incorporated as\cite{wet} $e^{\hat\mu}$ and $e^{-\hat\mu}$ multiplying $1\pm\gamma_4$ terms respectively in (\ref{eqn:Dwil}) so as to cancel $\hat\mu^2/a^2$ terms on the lattice\cite{karsch}. The energy density expression on the lattice has the same form as in (\ref{eqn:ed}) but with $r=\mu/T=\hat\mu N_T$ kept constant and  $h_4$ and $h_5$ changed to\cite{gatt}:
\begin{equation}
\label{eqn:edmu}
h_4=-\frac{a}{a_4}\sin(a_4 p_4 -i \hat\mu)~,~~
h_5=M-\sum _{j=1}^{3}(1-\cos a p_j)-\frac{a}{a_4}(1-\cos (a_4 p_4 -i \hat\mu))~.~
\end{equation}
The energy density in presence of $\hat\mu$ has been analytically shown to have the correct continuum limit\cite{gavai}.
 The two observables of interest here are the change in the energy density, $\Delta \epsilon (\mu,T) = \epsilon(\mu,T) -\epsilon(0,T)$ and the quark number susceptibility at $\hat\mu=0$ which in the continuum limit, are given by
\begin{equation}
\label{eqn:dem}
\frac{\Delta \epsilon(\mu,T)}{T^4}=\frac{r^4}{4\pi^2
}+\frac{r^2}{2}~~\&~~\chi(0,T)=\chi_{SB}=\frac{T^2}{3}
\end{equation}
As shown in figure 2, both these quantities also approach the continuum limit faster for the same optimum range of M. 
\begin{figure}
\begin{center}
 \includegraphics[scale=0.6]{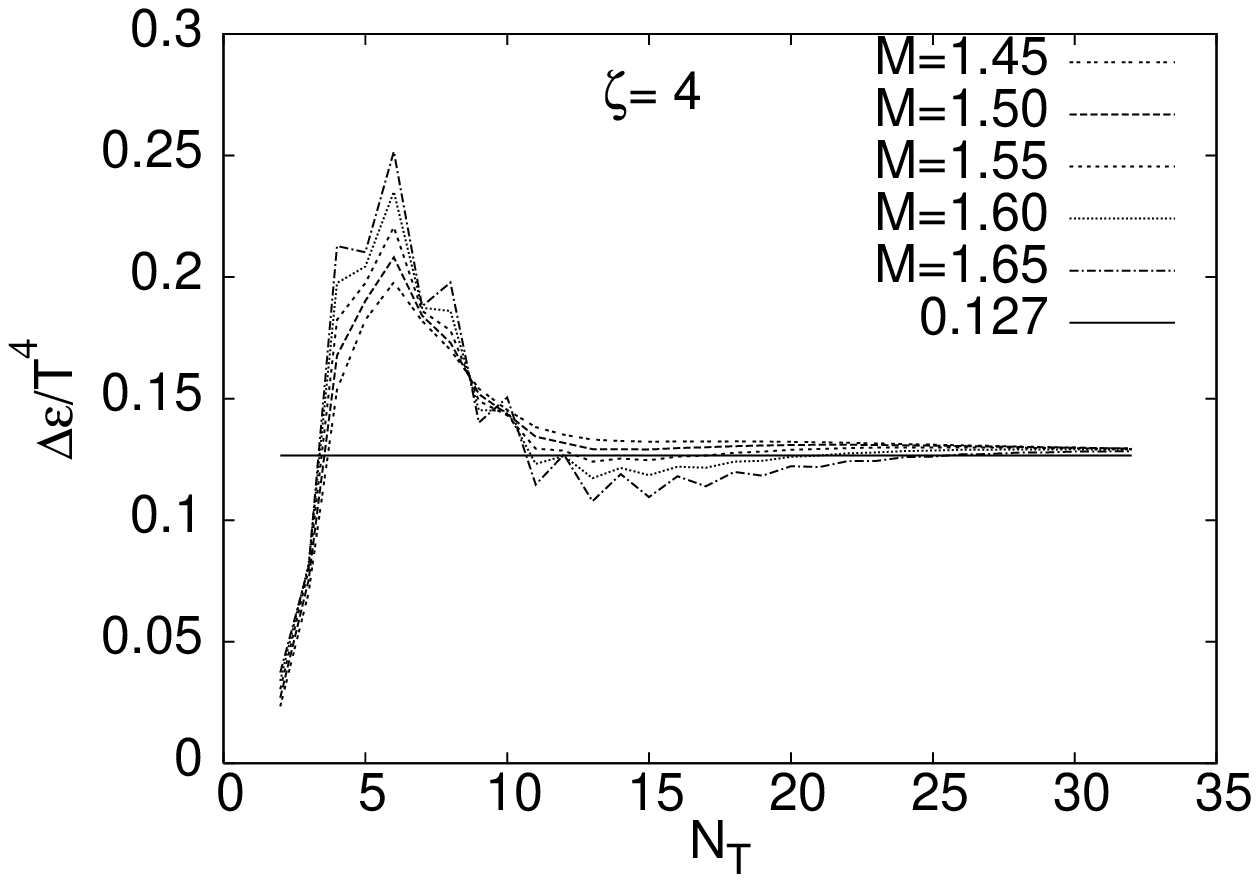}
 \includegraphics[scale=0.6]{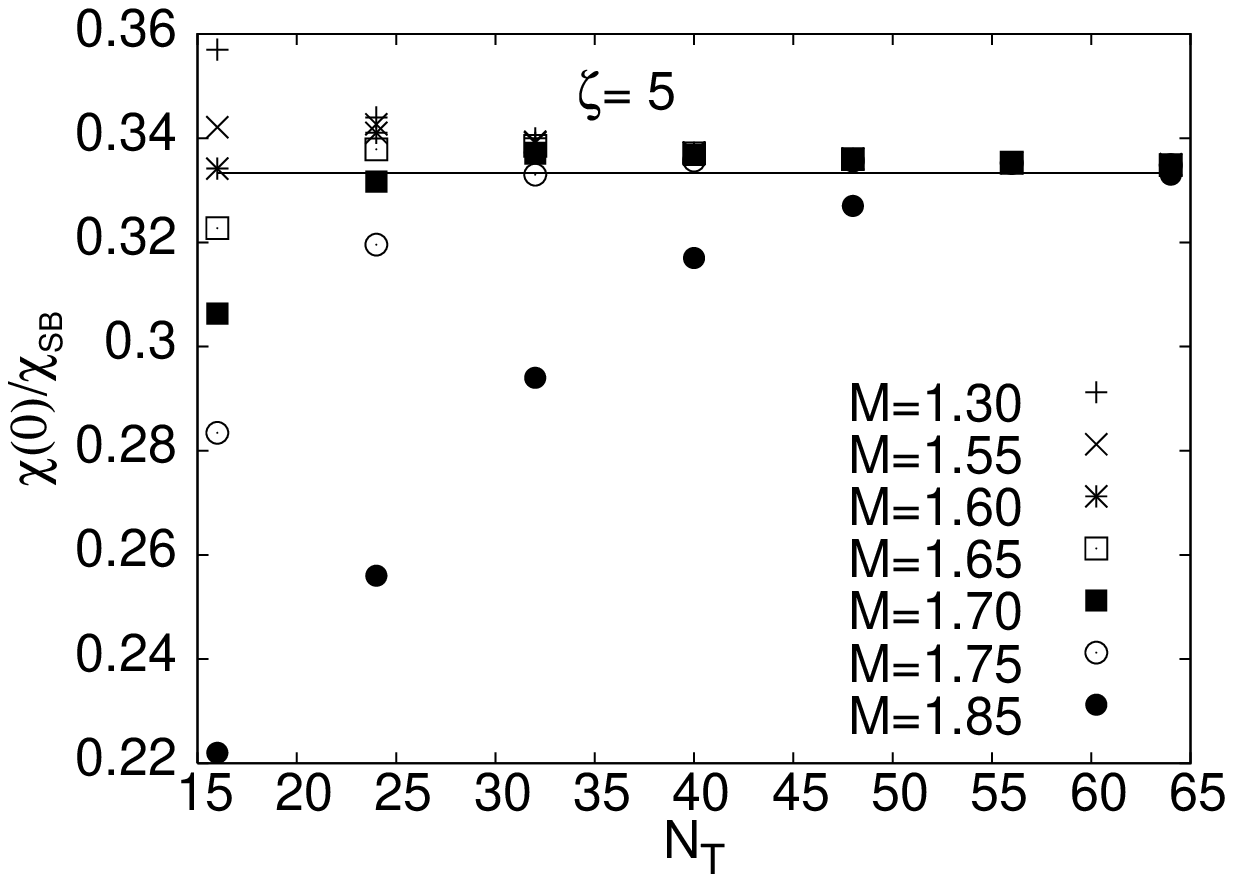}
\caption{(a)Comparison of the lattice $\Delta \epsilon/T^4$ with that for the
continuum for $r=0.5$. (b)The variation of $\chi(0)/\chi_{SB}$ vs $N_T$ for different M and 
$\zeta=5$.}
\label{DelE}
\end{center}
\end{figure}

\section{Conclusion}
The range $1.55<M<1.60$ is best suited for the computation of different thermodynamic quantities of free overlap and domain wall fermions on the lattice as the continuum results are obtained on the smallest possible lattice size. This range seem to be quite universal even in the presence of $\hat\mu$. The $L_5$ needed to obtain the overlap results starting from the domain wall
formalism is also minimum for this range of M. It is anticipated that this range would remain roughly same even in the presence of gauge fields and thus this work justifies the use of $M >1$ for faster computations even in full QCD with such fermions.

\section*{Acknowledgements}
It is a pleasure to thank our collaborator Debasish Banerjee. S.S would like to acknowledge the Council of Scientific and Industrial Research(CSIR) for financial support through the Shyama Prasad Mukherjee(SPM) fellowship.

\end{document}